\documentclass[a4paper,twoside]{article}
\newcommand{\affil}[1]{$^{\rm #1}$}
\usepackage[authoryear]{natbib}
\bibpunct{(}{)}{;}{a}{}{,}
\usepackage{graphicx,epsfig,url,amssymb}
\date{} %Please leave the date blank
\title{\large\bf\flushleft SPLASH: An interactive visualisation tool for Smoothed Particle Hydrodynamics simulations}
\author{\parbox{\textwidth}{\flushleft
\vspace{-0.5cm}
{\it Daniel J. Price\affil{A,B}}\\
\vspace{0.4cm}
{\small \affil{A}\,School of Physics, University of Exeter, Exeter EX4 4QL, UK}\\
{\small \affil{B}\,Email: dprice@astro.ex.ac.uk}}}
\begin{document}
\twocolumn[
\begin{changemargin}{.8cm}{.5cm}
\begin{minipage}{.9\textwidth}
\vspace{-1cm}
\maketitle
%
%
%%%%%%%%%%%%%     ABSTRACT    %%%%%%%%%%%%%
%Abstract of no more than 200 words here.
\small{\bf Abstract:}
This paper presents SPLASH,  a publicly available interactive visualisation tool for Smoothed Particle Hydrodynamics (SPH) simulations. Visualisation of SPH data is more complicated than for grid-based codes because the data is defined on a set of irregular points and therefore requires a mapping procedure to a two dimensional pixel array. This means that, in practise, many authors simply produce particle plots which offer a rather crude representation of the simulation output. Here we describe the techniques and algorithms which are utilised in SPLASH in order to provide the user with a fast, interactive and meaningful visualisation of one, two and three dimensional SPH results.

%%%%%%%%%%%%%     KEYWORDS    %%%%%%%%%%%%%
\medskip{\bf Keywords:} hydrodynamics -- methods: numerical
% Please write all keywords in lower case. PASA uses the
% standard list of subject headings adopted by The Astrophysical Journal
% and available from http://www.journals.uchicago.edu/ApJ/keywords_text.html.
% Keywords are separated by em-dashes, i.e. ---

%%%%%%%%DO NOT EDIT%%%%%%%%%%%%
\medskip
\medskip
\end{minipage}
\end{changemargin}
]
\small
%%%%%%%%EDIT FROM HERE%%%%%%%%%%%%

\section{Introduction}
%Please see the PASA Style Guide for help with correct layout for your manuscript.
%Examples of tables and figures are given below.
 Smoothed Particle Hydrodynamics \citep[for recent reviews see][]{monaghan05,price04} is a Lagrangian particle method for solving the equations of fluid dynamics. It has found widespread use in astrophysics due to the ability to simulate complicated three dimensional flow geometries and free surfaces with relative ease and the natural coupling with $N-$body techniques for self-gravitating problems. For example, SPH is used widely for simulations of cosmological structure formation \citep[e.g.][]{santabarbara,springel05}, for problems related to star \citep[e.g.][]{bbb03} and planet \citep[e.g.][]{mayeretal02} formation and in simulating astrophysical accretion discs\citep[e.g.][]{smithetal07} and stellar collisions  \citep[e.g.][]{fb05,dd06} and publicly-available SPH codes such as GADGET-2 by \citet{springel05} have found widespread application.
 
  However, visualisation of SPH data is not a straightforward process, since the data is defined on a set of moving points which follow the fluid motion and derivatives are evaluated by interpolation from neighbouring points weighted by a smoothing kernel. In practise many authors simply present particle plots which are a rather crude representation of the data. For example the widely used and publicly available Tipsy\footnote{http://www-hpcc.astro.washington.edu/tools/tipsy/tipsy.html} visualisation tool, though written for $N-$body simulations, is often used for SPH visualisation where the only procedure possible is to colour the particles according to the value of a scalar field such as density.
  
  A faithful visualisation of SPH data is much more complicated than for grid-based fluid codes since, for a smooth representation, a mapping procedure from the particles to a two dimensional array of pixels is required.  Using commercial visualisation packages (e.g. IDL) for this procedure is often inefficient because, for example, they require simply interpolating to a 3D grid first rather than mapping directly from the particles to the two dimensional pixel array required for a particular visualisation. Also, given that interpolation lies at the heart of SPH, consistency suggests use of the same interpolation algorithms as part of the visualisation procedure. Because fluid particles in SPH preserve their identity, there are also certain visualisation procedures which are possible which cannot be used in an Eulerian context, such as tracing the history of a portion of the flow via its component particles and tracking of particular objects. These aspects of SPH visualisation give strong motivation for a dedicated software tool designed to visualise SPH data using SPH algorithms. This paper presents the software design and algorithms implemented in exactly such a tool, which we have called ``SPLASH''. 
  
   SPLASH differs from other visualisation tools because it is designed specifically for SPH visualisation and works both interactively and non-interactively (see the discussion relating to the software design below). For example IFrIT\footnote{http://home.fnal.gov/$\sim$gnedin/IFRIT/} is a publicly-available tool written to visualise ionisation fronts in cosmological simulations (including those using particles) but allows only an interactive visualisation and lacks many of the features of SPLASH such as the ability to visualise in one, two and three dimensions, to select and hide particles and to track portions of the flow across multiple dump files. SPLASH allows plotting to both interactive and non-interactive devices allowing both a mouse-click driven visualisation as well as a ``pipeline'' mode for producing the same visualisation from a series of dump files (without the need for any kind of scripting). Similarly Splotch\footnote{http://dipastro.pd.astro.it/$\sim$cosmo/Splotch/} is a raytracing utility to visualise SPH simulations in a manner similar to the ``surface rendering'' technique implemented in SPLASH (see \S\ref{sec:renderplots}) but does not allow other visualisation techniques and does not have any interactive capabilities.

   Other publicly available tools such as Supermongo and Gnuplot implement primitive plotting functionality at a much lower level and would require a script of similar length to the SPLASH source code to achieve similar functionality in terms of visualising SPH data (equivalent to SPLASH's use of the PGPLOT library for actually plotting the results of the rendering operations). SPLASH can also be used to visualise remotely from the same location as the data is produced (e.g. on a remote supercomputing facility), installation on which is straightforward since the only requirement is a Fortran compiler which can also be used to compiler the PGPLOT libraries. Using a commercial package, this would not always be possible because it would require the remote facility to have the appropriate license (this in particular applies to IDL). Furthermore many visualisation tools require some form of scripting to achieve the desired functionality (in IDL's case, to the level of an entire programming language). Since SPLASH is specifically tailored to visualise SPH simulations with settings changed via a series of command-line based menus, no scripting is required even for complicated tasks such as producing a sequence of plots from multiple dump files (either interactively or non-interactively).

   The paper is organised as follows: In section \ref{sec:design} we discuss the basic requirements driving the software design and present the design in detail; in \S\ref{sec:plottypes} we discuss the basic methods for visualising SPH data and how these are incorporated into SPLASH and in \S\ref{sec:algorithms} we discuss the details of the interpolation algorithms implemented. Some additional features are described in \S\ref{sec:other} and the code's performance and memory usage are described in \S\ref{sec:perform}. A summary is given in \S\ref{sec:summary}.

\section{Software design} 
\label{sec:design}
 
\begin{figure*}
\begin{center}
\epsfig{file=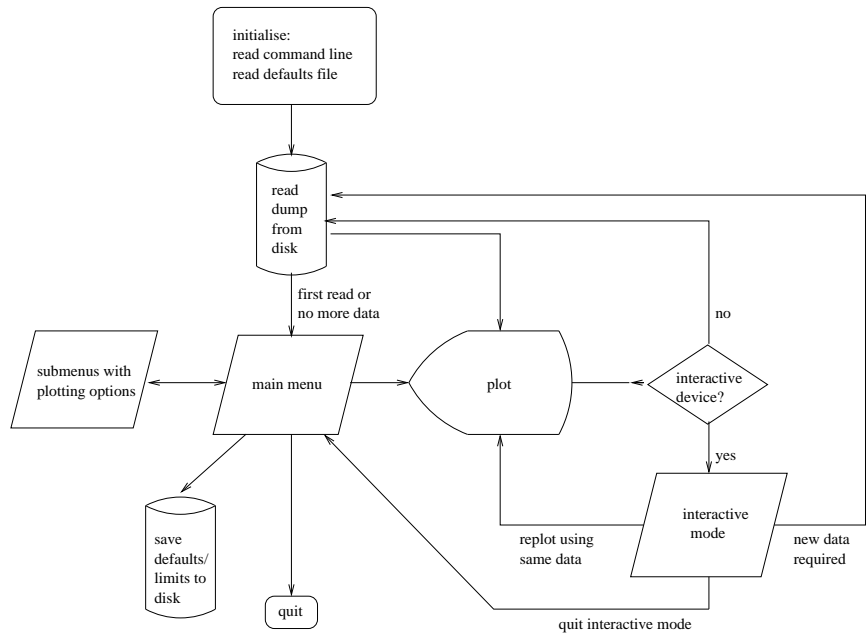,width=0.7\textwidth}
\caption{Basic software design}
\label{fig:flowchart}
\end{center}
\end{figure*}

   The basic requirements I set for an SPH visualisation tool (based largely on my own experience of performing SPH simulations) were as follows:
\begin{enumerate}
\item capable of producing sufficiently annotated, appropriately labelled figures suitable for inclusion in research papers
\item capable of producing a sequence of images for making animations
\item capable of reading data directly from binary code dumps from users' SPH codes
\item visualisation of SPH data in 1, 2 and 3 dimensions
\item algorithms should be consistent with the basic SPH method
\item should be easy to apply the same visualisation to different dump files (either interactively or non-interactively)
\item visualisation of both scalar and vector fields defined on the particles
\item visualisation should be interactive so the user can rapidly understand the data and find the best representation
\item remote visualisation capability -- since simulation data is often produced remotely on supercomputing facilities from which data transfer is awkward and time-consuming
\item written in a programming language familiar to users
\end{enumerate}

 SPLASH is a program designed to meet these basic visualisation requirements in the most efficient manner possible. Each of the above requirements have strongly constrained the software design. For example the requirement that the visualisation be interactive means that simple but inefficient procedures such as interpolating from the particles to a 3D grid before using standard grid-based visualisation techniques cannot be utilised. 

  The basic software design which achieves all of the above is outlined in Figure~\ref{fig:flowchart}. The code (written in Fortran~90) is built around a command-line menu structure (designed so as to meet the requirement for remote visualisation) with the actual plotting performed via the PGPLOT graphics subroutine library\footnote{http://www.astro.caltech.edu/$\sim$tjp/pgplot} (thus satisfying the requirements for production of figures for papers -- via the postscript device drivers; for movies -- via bitmap device drivers such as PNG and GIF; for interactivity -- via interactive devices such as the X-windows driver). The use of a graphics library not only facilitates the easy reproduction of the same plots on different devices but also means that SPLASH can be focussed on the data input and manipulation side of the visualisation procedure rather than the implementation of primitive plotting functionality.

Plot settings are changed either non-interactively via a series of sub-menus accessed on the command line from the main menu; or interactively using the mouse and/or pressing particular keystrokes with the cursor in the plotting window (this is the ``interactive mode'' indicated in Figure~\ref{fig:flowchart}). 

Rather than requiring the user to convert data to an intermediate format (e.g. ascii files), data is read directly from the binary code dump files -- this is a crucial requirement for rapid visualisation and makes for significantly reduced disk space requirements (since no intermediate storage is required), which can be a major constraint on many systems for simulations involving $\gtrsim 10^{6}$ particles. The filenames are read from the command line, making it easy to read all files from a simulation by using wildcards (e.g. ``\verb+splash dump*+''). Read routines are supplied for several widely used SPH codes (e.g. GADGET, \citealt{springel05}; VINE, \citealt{wetzstein07}; and Matthew Bate's SPH code, \citealt{bate95}). Optionally, a further set of derived quantities can be calculated from the data read. For a typical SPH data set this would include the radius, the magnitude of all vector quantities and the entropy. These quantities appear as ``extra columns'' as if they had been read from the dump file.

 The first file listed on the command line is read on entry (see Figure~\ref{fig:flowchart}) and this determines the basic parameters used for the visualisation such as the number of data columns, column labels and unit settings where appropriate. The data is then listed by column in the main menu, where the column number corresponds to the position of a variable in the data read. This means that any two parameters can be plotted against one another (for example density vs x would be plotted by typing 6 for the y-axis and 1 for the x-axis assuming column 1 contains the particle x position and column 6 contains the density). Where two of these columns correspond to particle coordinates we refer to this as a ``coordinate plot'' which (provided particle masses, density and smoothing lengths have been read from the dump file) can be plotted either as particle plots or with a third quantity ``rendered'' to a pixel array. Thus, for example, a plot of density in a two dimensional domain is a plot of y vs x with density rendered. If vector quantities are present in the data (specified in the data read corresponding to that particular data format) a fourth quantity can also be plotted over the rendered plot in the form of an arrow plot. In 3D these plots can either be projection (using all particles) or cross section plots (using only particles contributing to a slice positioned in the third coordinate direction). Similarly two dimensional ``rendered'' plots are either plots using all of the particles or line plots tracing an oblique cross section through the computational domain. The interpolation procedure used to map from the particle data to a rendered image are described below and the algorithms are presented in \S\ref{sec:algorithms}.
 
 The plotting is directed to a particular device via a PGPLOT prompt. For interactive devices, the program then enters ``interactive mode'', where the user can manipulate the data interactively either using the mouse (to zoom, change colour bar limits, select and colour particles and move legend positions) or via keystrokes pressed in the plot window, giving access to a wide range of options such as: rotating the particles, moving the 3D observer, adapting plot limits, plotting smoothing circles, labelling particles, changing the colour scheme, adjusting the length of arrows on vector plots, setting up animation sequences, finding the gradient of a line and (most importantly) moving forwards and backwards through timesteps. For example pressing the space bar moves forwards to the next dump file, whereupon the same plot is repeated (and repeatedly pressing the spacebar produces a crude `animation' dependent of course, on the speed at which data can be read from disk and plotted). This is indicated by the loop in Figure~\ref{fig:flowchart} which proceeds from interactive mode via the data read back to the ``plot'' step and finally returning to interactive mode. Where no data read is required the plot is simply re-plotted with the changed settings (perhaps recalculating the interpolation to pixels where necessary).
 
  A key feature facilitating the easy production of animations is that, when plotting is directed to a non-interactive device, the plotting cycles automatically through all of the dump files on the command line. This is indicated by the loop in Figure~\ref{fig:flowchart} proceeding from the ``plot'' step back to the data read (if the device is non-interactive) and returning plot the same figure for the next dump file with settings unchanged.

 The settings for a particular plot can be saved to disk by pressing `s' from the main menu (see Figure~\ref{fig:flowchart}. This saves a file in the current working directory containing (in Fortran 90 NAMELIST format) all of the current plot settings. This file is then read automatically on the next invocation of SPLASH such that plot settings can be restored. A ``full save'' (implemented by pressing `S' from the main menu) saves both the plot settings and the current minimum and maximum limits set for each column (in a simple two-column ascii file), so that \emph{exactly} the same plot can be reproduced on the next invocation of SPLASH. Additional files are also saved where physical units have been applied to the data columns or animation sequences have been set.
 
  The plot settings are structured into Fortran~90 modules which contain the parameters which may be changed via a particular submenu together with the subroutine implementing the submenu itself. Each settings module contains it's own namelist for those parameters which should be saved to disk. Thus the `save' operation simply saves all of the namelists in order into a single file. This structure means that, for the programmer, it is a straightforward task to add additional menu options affecting particular plotting functions (e.g. settings related to vector plots are changed in a ``vector plot options'' submenu and both the settings and the submenu are contained in the same Fortran~90 module. This module is then USE-d only in the subroutines which implement the plotting of vector plots, so any parameters changed via options in the vector submenu will be automatically available near where they will be used to make plotting decisions and automatically saved to the defaults file provided they have been added to the namelist).

\section{Plot types}
\label{sec:plottypes}
\begin{figure}
\begin{center}
\epsfig{file=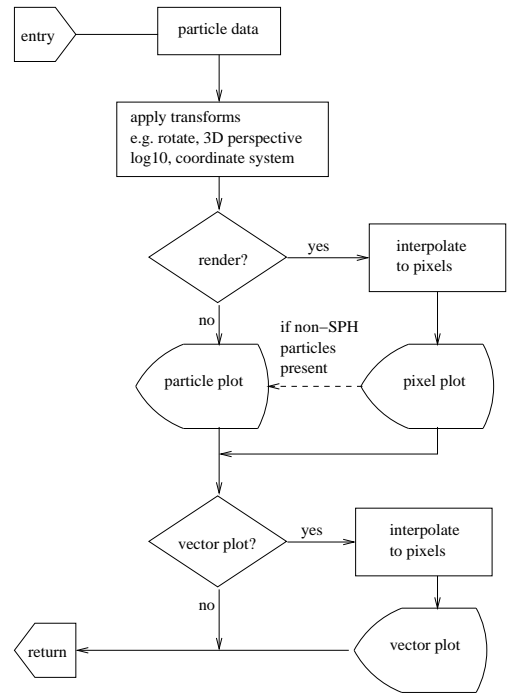,width=0.85\columnwidth}
\caption{Plotting pipeline}
\label{fig:plotpipeline}
\end{center}
\end{figure}

 The  ``central engine'' of the visualisation procedure is encapsulated in the ``plot'' step in Figure~\ref{fig:flowchart}. An expanded outline of this step is shown in Figure~\ref{fig:plotpipeline}. There are essentially two types of plots: particle plots or rendered plots, where a further rendered plot of vector arrows can be plotted on top of either of these. The procedure for each of these is described in turn below.  Note that transformations such as log, rotation, 3D perspective and change of co-ordinate systems are applied to the particle data prior to calling any interpolation routines. 

\subsection{Particle plots}
\begin{figure}[h!]
\begin{flushleft}
\includegraphics[width=0.8\columnwidth,angle=270]{orstang_part.ps}
\includegraphics[height=0.8\columnwidth]{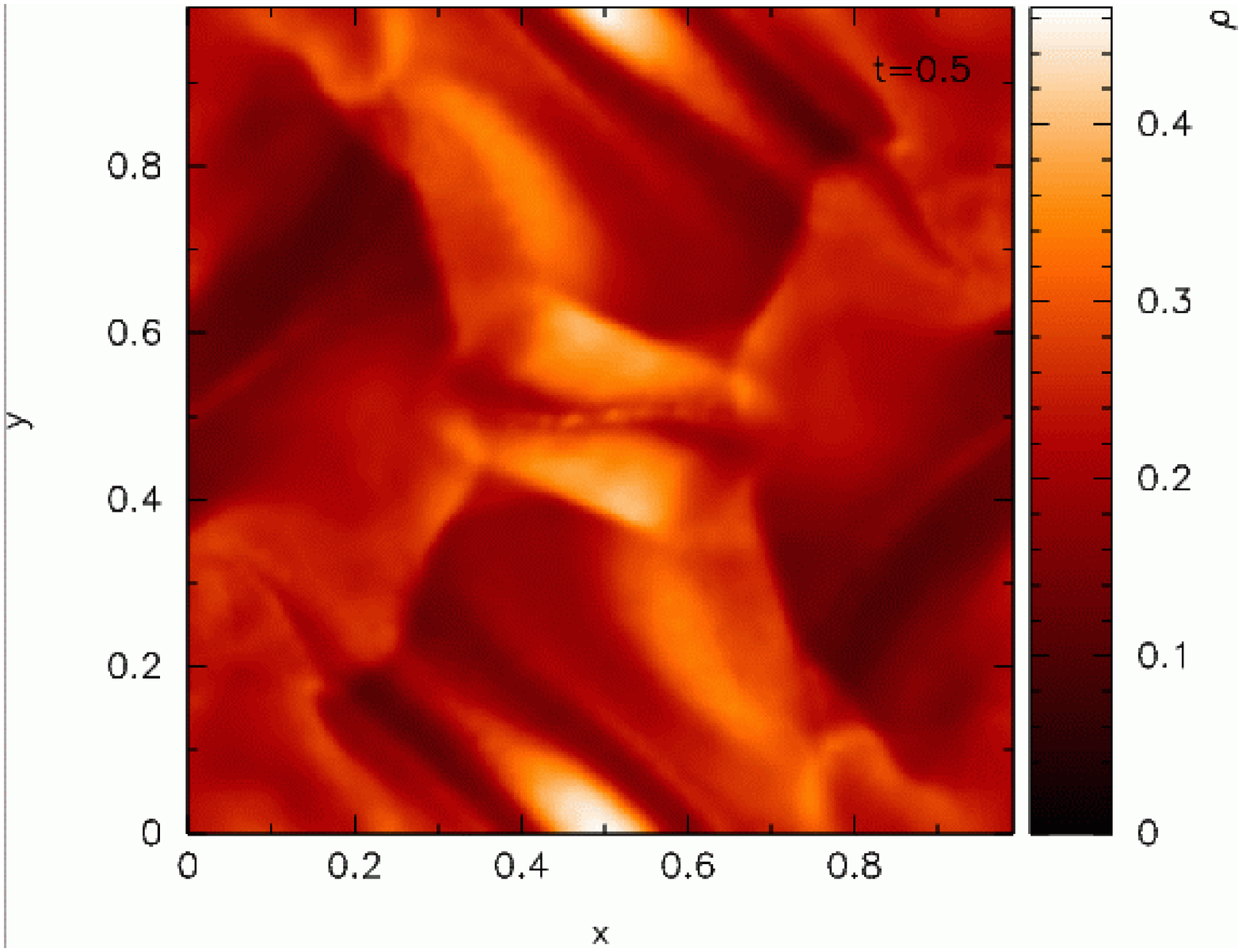}
\includegraphics[height=0.8\columnwidth]{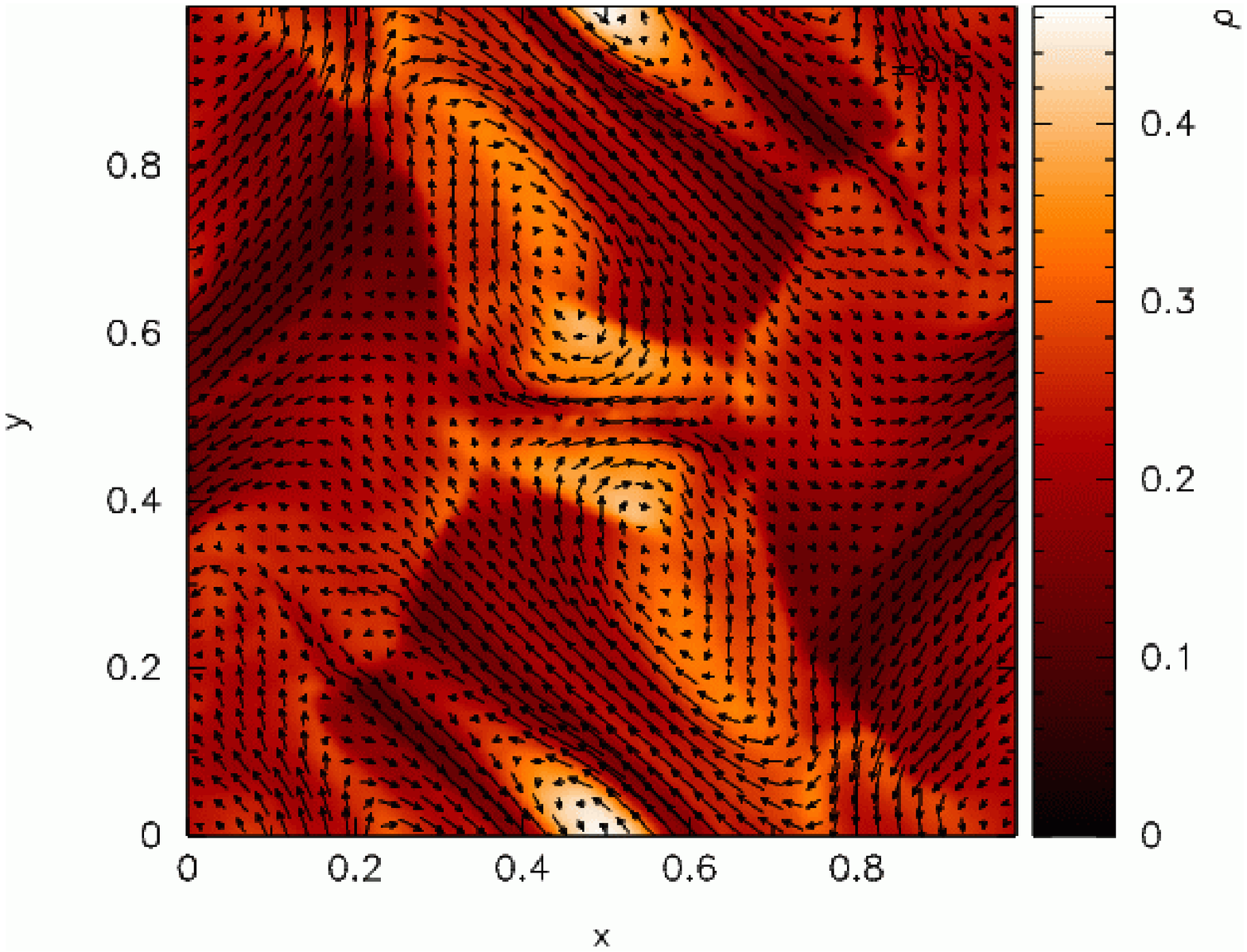}
\caption{Top panel: A simple particle plot produced from a 2D simulation simply by plotting all of the particle positions. Middle panel: The same data but plotted with data rendered to a pixel array instead of plotting particles. Bottom panel: with a vector plot additionally overlaid (in this case showing the magnetic field in the simulation).}
\label{fig:orstang}
\end{flushleft}
\end{figure}

 If rendering is not being used (ie. the plot is not a coordinate plot or no third quantity has been selected), the plot can simply proceed by plotting the particle positions directly on the plotting device (Figure~\ref{fig:plotpipeline}), using markers which can be chosen dependent on the particle types (set via the submenus accessed from the main menu -- see Figure~\ref{fig:flowchart}). A simple particle plot example is shown in the top panel of Figure~\ref{fig:orstang}. Particle colours can be changed in a variety of ways. For example, selecting particles with the mouse and pressing keys 1-9 whilst in interactive mode colours the selected particles with colours corresponding to the respective PGPLOT colour indices. Since colours allocated to particles are retained in all subsequent plots, this can be used to select ranges of a particular parameter (e.g. by selecting particles on a density vs x plot) with the colours still appearing on a different plot (e.g. a coordinate plot of y vx x). Similarly particles can be coloured using data from a dump corresponding to the initial conditions and, provided particles retain their identity between dumps, the same particles will still appear coloured when plotted in subsequent dumps.
 
 Particles can also be coloured according to the value of a particular quantity by setting an option which renders via particle colours instead of interpolating to pixels, although the latter method (see below) is strongly preferred as a method of visualisation. However there are circumstances where it may be desirable to see the actual values of a quantity on the particles themselves. 
 
  Where the `cross section' option has been set from the menu, particles are plotted in a thin slice of finite (although user-adjustable) thickness around the user-defined cross section slice position.

\subsection{Rendered plots}
\label{sec:renderplots}

 ``Rendered'' or ``pixel'' plots proceed in a similar manner to particle plots but with an intermediate step where the particle data is interpolated to the two-dimensional pixel array corresponding to the viewing surface (Figure~\ref{fig:plotpipeline}). In 3D rendered plots are either projections (integration along the line of sight), cross section slices or surface rendered plots (see \S\ref{sec:surface}). In 2D the plot can either be a projection (a straight interpolation to a 2D pixel array) or a cross section (a 1D line plot drawn arbitrarily through the 2D domain). Rendered plots do not apply in the case of 1D data.  An example of a 2D rendered plot is shown in the middle panel of Figure~\ref{fig:orstang}, where the same data shown in the particle plot (top panel) has been used. Note the striking difference between the visualisation using pixels compared to the raw particle plot (this kind of plot is often used as a representation of the density field). One of the goals of SPLASH is to make visualisation of SPH data in this manner a straightforward task for the user.
 
\begin{figure}
\begin{flushleft}
\includegraphics[width=\columnwidth]{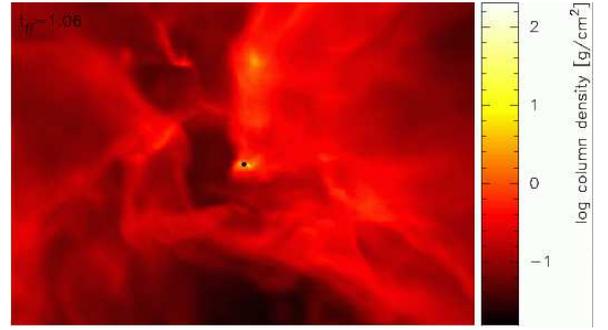}
\caption{Visualisation of a large scale star cluster formation calculation (Price \& Bate 2007) via a 3D rendered plot showing the density integrated through the $z$-direction.}
\label{fig:mcluster}
\end{flushleft}
\end{figure}

  A slight complication here is that often simulations contain particles of multiple types, some SPH (e.g. different types of gas particle) and some non-SPH (e.g. sink or $N-$body particles). In this case the interpolation is performed using all of the available SPH particles, provided plotting of that type has been turned on via the submenu options. Particles of non-SPH types can optionally be plotted on top of rendered plots (e.g. sink particles appear on top of a rendered plot of gas density) -- this is indicated by the dashed pathway in Figure~\ref{fig:plotpipeline}. An example of a three dimensional rendered projection (ie. showing in this case column density) of a large scale star formation calculation (similar to that described in \citealt{bbb03}) is shown in Figure~\ref{fig:mcluster}, where additionally a sink particle has been plotted over the rendered array.

\begin{figure}[h]
\begin{center}
\includegraphics[width=\columnwidth]{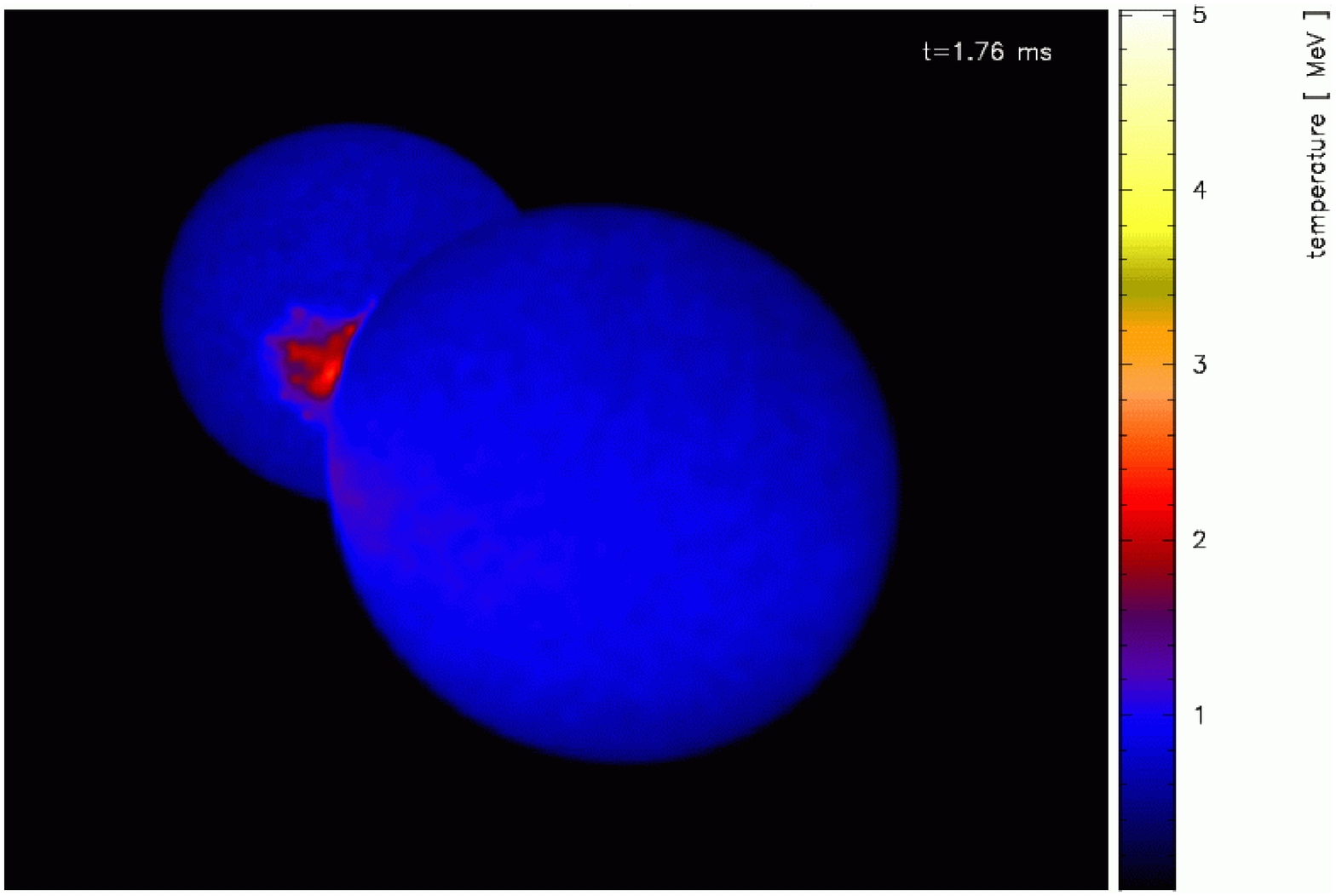}
\includegraphics[width=\columnwidth]{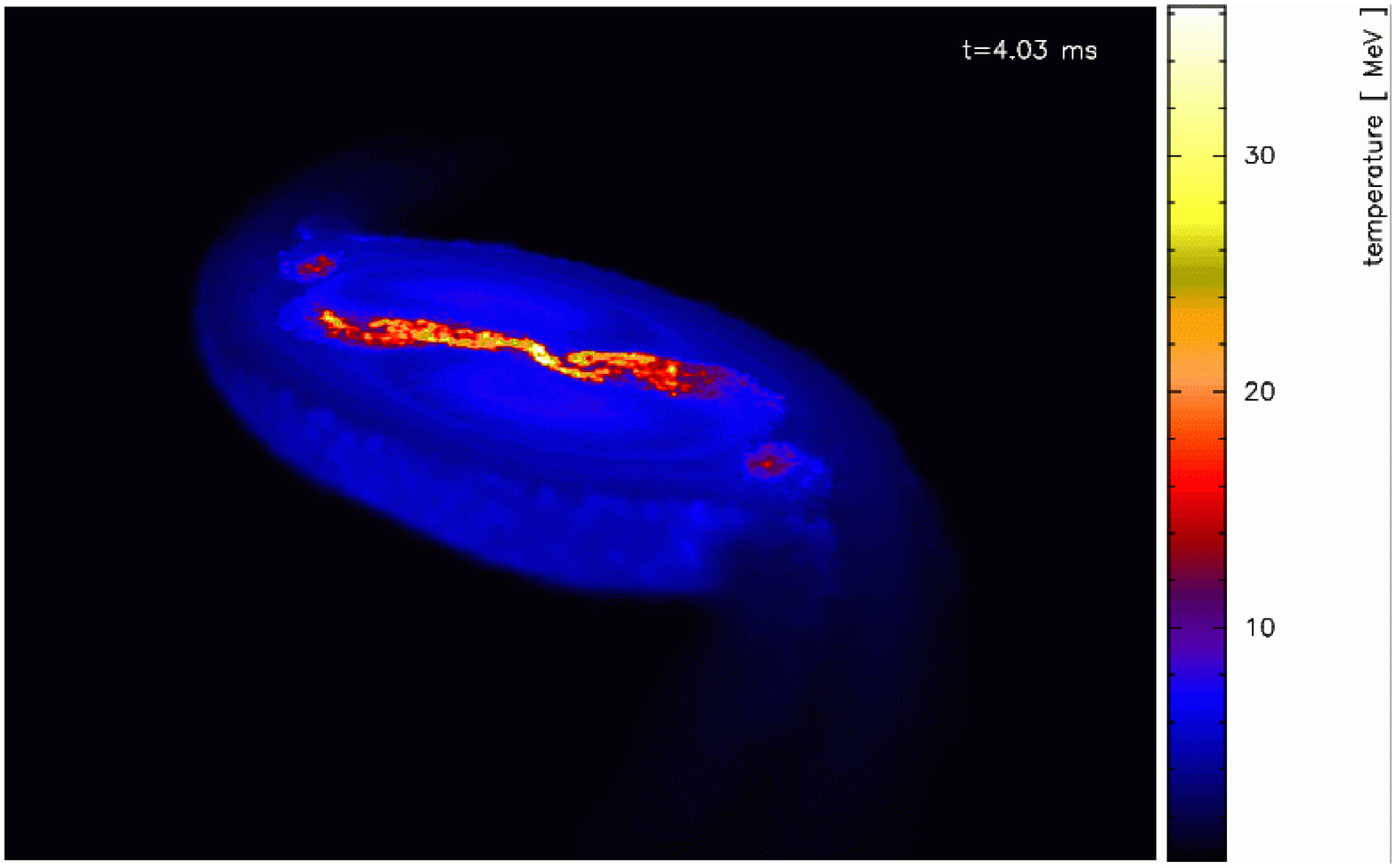}
\caption{Visualisation of a neutron star merger calculation \citep{pr06} via a 3D surface rendered plot showing the temperature on a `surface of last scattering'. The top panel shows results near the start of the simulation using all of the particles, whereas in the bottom panel only particles below the mid-plane have been used in the interpolation, producing a `cut-away' effect.}
\label{fig:nsns}
\end{center}
\end{figure}

 One further type of rendering is available in SPLASH for three dimensional data, which we refer to as ``surface rendering'' (the algorithm is described in \S\ref{sec:surface}, below). This type of rendering provides an ``optically thick'' view of the particles (as opposed to the ``optically thin'' view provided by the column integrated rendering), showing the value of a particular parameter on the ``surface of last scattering'', determined by a user-defined opacity which is proportional to the particle density. Generally this type of visualisation works best for simulations where there is a well-defined surface and/or the range of densities in the simulation is not too high. An example is shown in Figure~\ref{fig:nsns} showing gas temperature during the merger of two neutron stars similar to those described in \citet{pr06}. The top panel shows a surface rendering near the start of the simulation where all of the particles have been used in the interpolation. The bottom panel shows a similar plot but where only particles below the midplane have been used in the calculation, giving a `cut-away' effect.
 
\subsection{Vector plots}
 Vector quantities are visualised using arrow plots, although more advanced visualisations may be possible in future. Whilst in principle an arrow could be plotted for each SPH particle with length proportional to the value of the vector on that particular particle, these type of plots quickly become cluttered when large numbers of particles are used in the simulation. Thus vector plots in SPLASH are implemented by first interpolating each component of the vector quantity to the two-dimensional pixel array corresponding to the viewing surface, where in 3D the plot can be an integration of each component along the line of sight or where vector arrows are plotted in a cross section slice (depending on whether cross sections or projections have been selected in the menu options, also affecting rendered plots). 
 
  An example of a vector plot is shown in the lower panel of Figure~\ref{fig:orstang} where the arrows are shown overlaid on the rendered plot of density (otherwise identical to the middle panel). A preliminary feature has also been implemented whereby streamlines can be calculated for the interpolated vector field and plotted as contours (instead of plotting arrows). As presently implemented (see \S\ref{sec:streamlines}) this works quite well when the vector field is smooth but gives poor results where the field has strong gradients. A strongly desirable feature for future implementation would be an algorithm for tracing three dimensional field lines through SPH particle data.

\section{Interpolation algorithms}
\label{sec:algorithms}

\subsection{SPH interpolation}
 The heart of the SPH method (see e.g. \citealt{monaghan92,price04,monaghan05} for reviews) is the following identity for an arbitrary function $A({\bf r})$ defined on spatial coordinates ${\bf r}$:
\begin{equation}
A({\bf r}) = \int A({\bf r}') \delta(\vert {\bf r} - {\bf r}' \vert) d{\bf r}',
\end{equation}
where $\delta$ is the Dirac delta function. This integral is approximated in SPH by replacing the delta function with a smooth function $W$ with finite characteristic width $h$ which reduces to a delta function in the limit $h\to 0$, giving the SPH `integral interpolant' in the form
\begin{equation}
A({\bf r}) = \int A({\bf r}') W(\vert {\bf r} - {\bf r}' \vert, h) d{\bf r}' + \mathcal{O}(h^{2}),
\end{equation}
where the error in the representation of $A$ is of order $h^{2}$ provided the kernel function $W$ is even and the kernel function is normalised such that the volume integral of the kernel is unity. This integral is discretised onto the particles by replacing the integral with a summation over neighbouring particles and replacing the mass element $\rho d{\bf r}'$ with the neighbouring particle mass $m$, ie.
\begin{equation}
A({\bf r}) \approx \sum_{j=1}^{N} \frac{m_{j}}{\rho_{j}} A_{j} W(\vert {\bf r} - {\bf r}_{j}\vert, h).
\label{eq:suminterp}
\end{equation}
where the subscript $j$ refers to a quantity defined on particle $j$. The expression given above is the SPH `summation interpolant', forming the basis of the SPH approach and therefore the basis of the interpolation algorithms used in SPLASH for SPH visualisation. A normalised version of this interpolant is achieved by dividing the result by the interpolation of unity, given by
\begin{equation}
1 \approx \sum_{j=1}^{N} \frac{m_{j}}{\rho_{j}} W(\vert {\bf r} - {\bf r}_{j}\vert, h).
\label{eq:norm}
\end{equation}

\begin{figure}[h!]
\begin{center}
\includegraphics[width=\columnwidth]{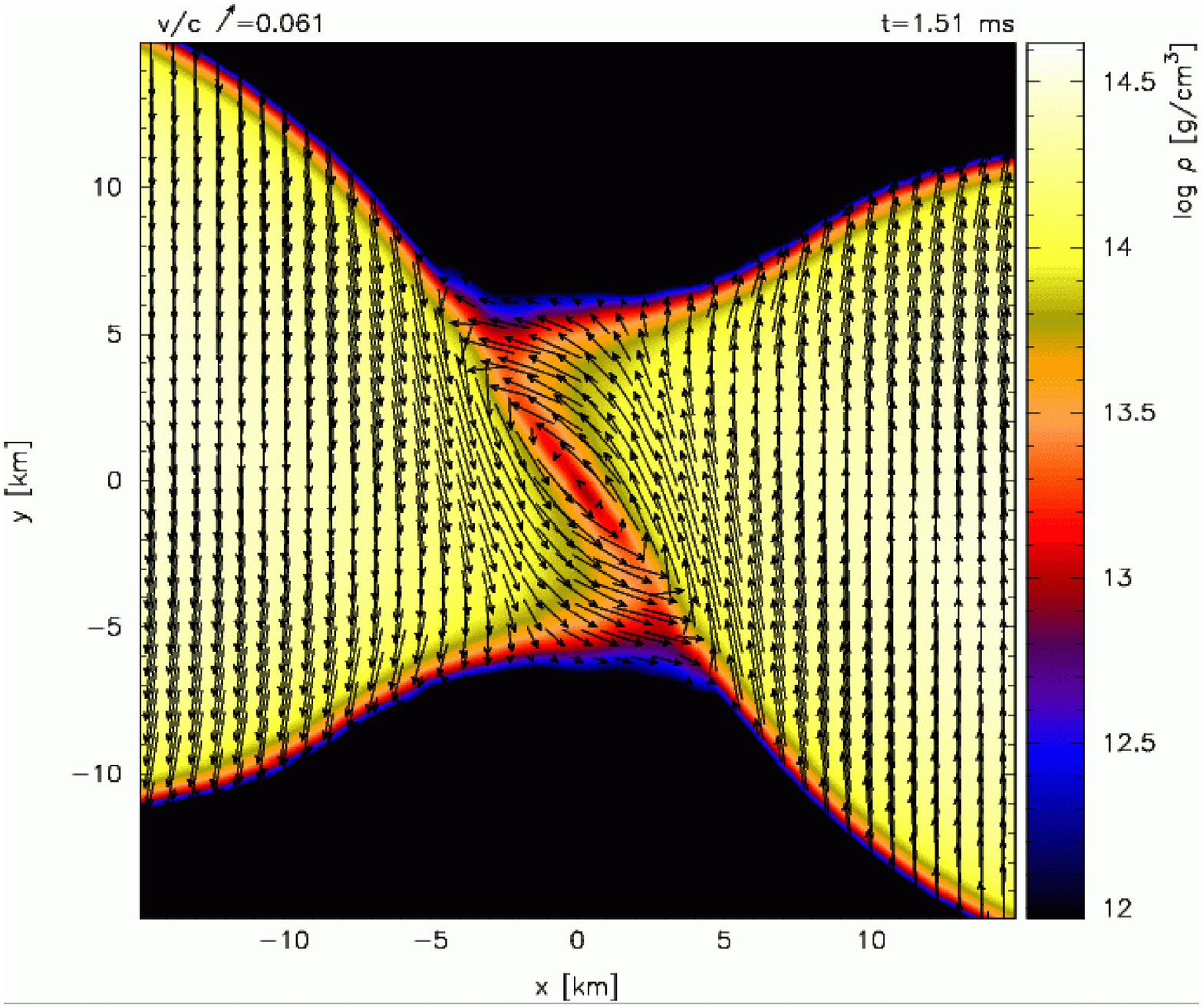}
\includegraphics[width=\columnwidth]{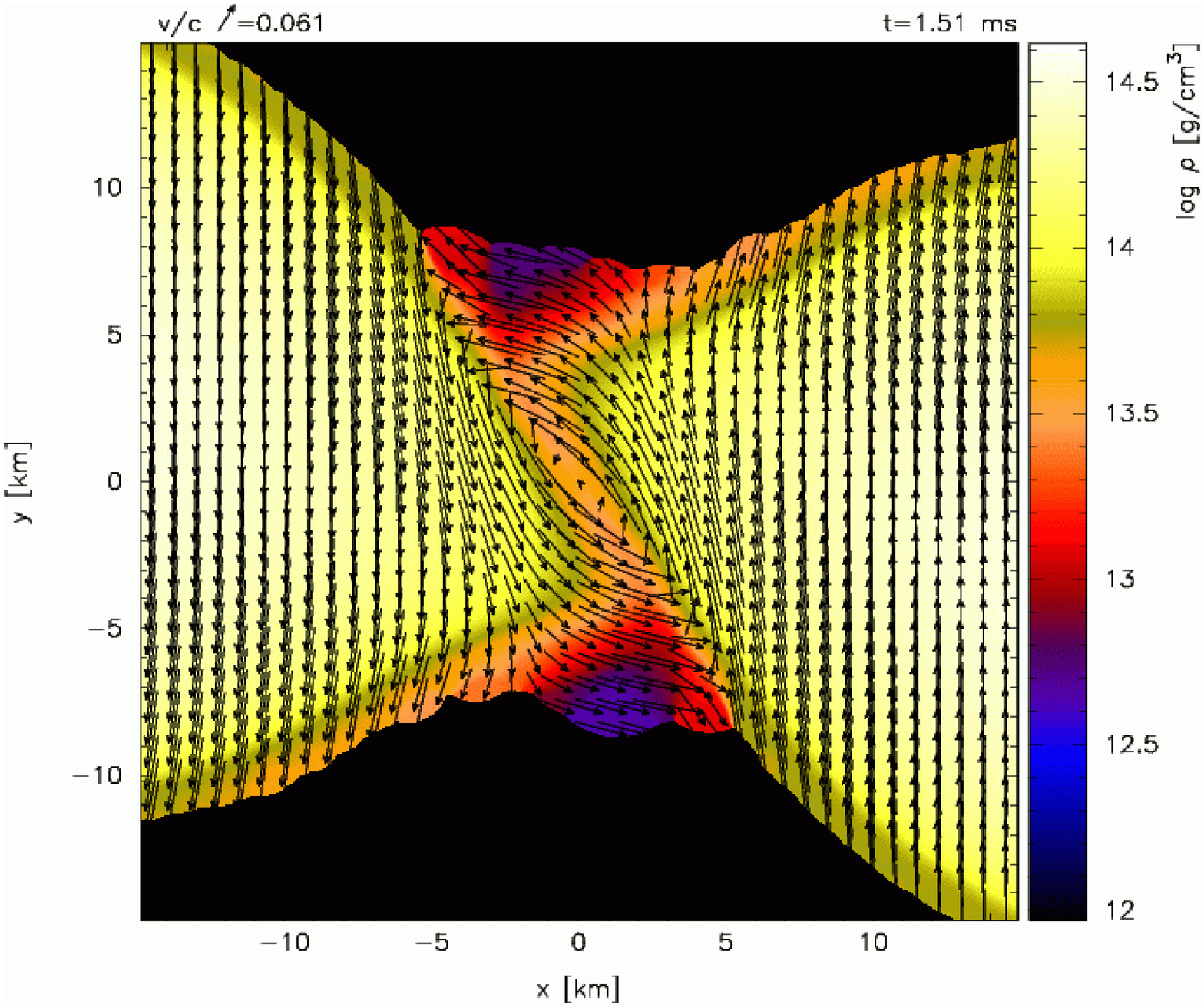}
\caption{Cross section slice of density (and velocity arrows) in a neutron star merger calculation \citep{pr06} showing the difference between non-normalised (top) and normalised (bottom) interpolation. Normalised interpolation is turned off by default as it produces spurious effects due to individual particles at free surfaces (bottom panel).}
\label{fig:normalisation}
\end{center}
\end{figure}

 Many different forms are possible for the smoothing kernel $W$, but the most commonly used is the cubic spline kernel (see \citealt{monaghan92}):
 \begin{equation}
W(r,h) = \frac{\sigma}{h^\nu}\left\{ \begin{array}{ll}
1 - \frac{3}{2}q^2 + \frac{3}{4}q^3, & 0 \le q < 1; \\
\frac{1}{4}(2-q)^3, & 1 \le q < 2; \\
0 & q \ge 2 \end{array} \right. \label{eq:cubicspline}
\end{equation}
where $q = \vert {\bf r}_a - {\bf r}_b \vert / h$, $\nu$ is the number of spatial
dimensions and the normalisation constant $\sigma_{\nu}$ is given by $\sigma_{1} = 2/3$, $\sigma_{2} = 10/(7\pi)$ and $\sigma_{3} = 1/\pi$. This kernel satisfies the basic requirements that it is Gaussian-like and has smooth first derivatives which tend smoothly to zero as $q \to 2$ and is zero beyond $q=2$.
The quantity $h$ is the smoothing length, which in most astrophysical applications is a spatially variable quantity set in such a way as either to fix (either exactly or approximately) the number of nearest neighbours \citep{hk89,benzetal90}, or via an analytic relation to the (number) density \citep{sh02,monaghan02,pm07}.

 By default the interpolations used in SPLASH are non-normalised. The reason for this is that, at a free surface the normalised interpolation (that is, using Eqn. \ref{eq:suminterp} and dividing the result by Eqn.~\ref{eq:norm}) looks odd, whereas an interpolation using Eqn. (\ref{eq:suminterp})  falls away smoothly. An example is shown in Figure~\ref{fig:normalisation} which shows a cross section slice of density from a three dimensional neutron star merger calculation \citep{pr06}. The top panel shows the results using a non-normalised interpolation whereas the bottom panel shows the results when the interpolated array is normalised (by dividing by the interpolation of unity). The normalised interpolation performs poorly at the edges, where the effects of individual particle smoothing spheres are visible. However, using a normalised interpolation improves the accuracy of volume rendered quantities on the pixels by removing effects due to the particle distribution. Thus it is recommended that a normalised interpolation should always be used if there are no free surfaces.

 To avoid round-off error in interpolation calculations (done in single precision), we write the summation interpolant in the simpler form:
\begin{equation}
A({\bf r}) \approx \sum_{j=1}^{N} w_{j} A_{j} \mathcal{W}(r/h).
\label{eq:interp}
\end{equation}
where $w_{j}$ is the dimensionless weight given by
\begin{equation}
w_{j} \equiv \frac{m_{j}}{\rho_{j} h_{j}^{\nu}},
\label{eq:weight}
\end{equation}
where $\nu$ is the number of spatial dimensions and $\mathcal{W}$ refers to the dimensionless part of the kernel function, such that
\begin{equation}
W(\vert {\bf r} - {\bf r}_{j}\vert, h) = \frac{1}{h^{\nu}} \mathcal{W}(r/h),
\end{equation}
(ie. we have incorporated the $1/h^{\nu}$ part of the usual kernel definition into the weight). 

 With this definition a normalised interpolation is given by
\begin{equation}
A({\bf r}) \approx \frac{\sum_{j=1}^{N} w_{j} A_{j} \mathcal{W}(r/h) }{\sum_{j=1}^{N} w_{j} \mathcal{W}(r/h) }
\end{equation}

 As an interesting aside, it is worth noting that the usual formula for varying the smoothing length in SPH codes is given by
\begin{equation}
h = \eta \left( \frac{m}{\rho} \right)^{1/\nu},
\end{equation}
where $\eta$ is a constant and $\nu$ refers to the number of spatial dimensions. Enforcing this relation rigourously (e.g. as in \citealt{sh02,pm04b,pm07}) thus corresponds to using constant weights (Equation \ref{eq:weight}) in the interpolation with the value related to the parameter $\eta$. Thus strictly, only knowledge of the (constant) weight value and the smoothing length is required for interpolation of any quantity in these codes.

\subsection{Rendering of 2D data}
\begin{figure}
\begin{center}
\includegraphics[width=0.8\columnwidth]{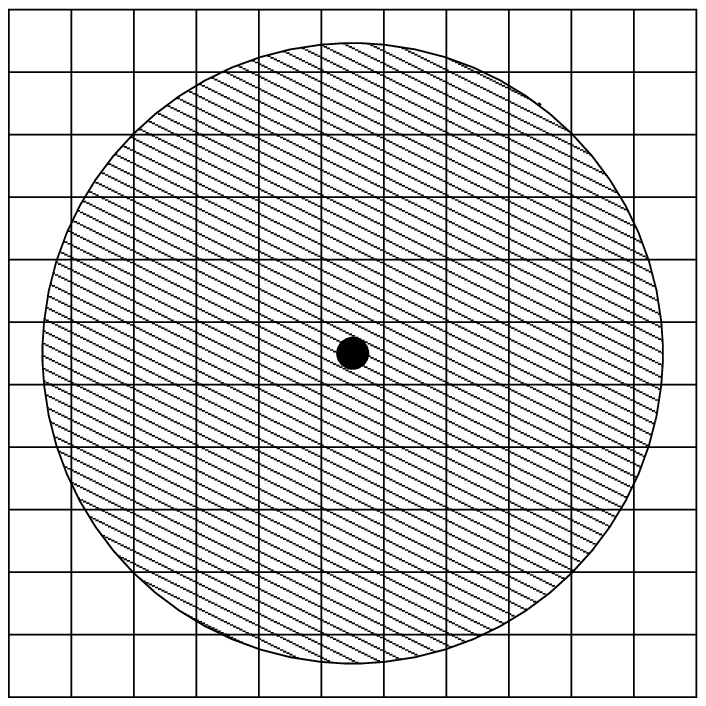}
\caption{Interpolation of 2D data: For each particle we perform a loop over the pixels (in $x$ and $y$) to which it contributes, adding the contribution from that particle to the pixel array.}
\label{fig:2Dinterp}
\end{center}
\end{figure}

\subsubsection{Interpolation to pixels}
 Rendering of 2D data involves a straightforward application of Eqn. (\ref{eq:interp}) to the interpolation of data from the particles to a two dimensional grid of pixels. Thus we have
\begin{equation}
A(x,y)  =  \sum_j w_{j} A_{j} \mathcal{W}(r / h), \label{eq:2dinterp} 
\end{equation}
where
\begin{equation}
r = \sqrt{(x - x_j)^{2} + (y -y_j)^{2}}, 
\end{equation}
the summation is over contributing particles and we take the smoothing length as
\begin{equation}
h = \mathrm{max}(h_{j},\Delta/2),
\end{equation}
that is, the maximum of the particle smoothing length and half of the pixel width (the latter thus being used generally only when few pixels are used in the interpolated plot). The interpolation is performed as a ``scatter'' operation from the particles, that is, for each particle $b$, we find the range of pixels to which the particle should contribute (in both $x$ and $y$) and add the contribution from particle $b$ to all of those pixels. Note that this is much more efficient than attempting to perform the summation over particles in Equation (\ref{eq:2dinterp}) for every pixel. The procedure is illustrated in Figure~\ref{fig:2Dinterp} and examples of 2D interpolation are shown in Figure~\ref{fig:orstang}.

\subsubsection{Cross sections of 2D data}
 The cross-sectioning algorithm for 2D data (giving a 1D line) is completely general and can be used for arbitrary oblique (or straight) cross sections. The cross
section is defined by two points (x1,y1) and (x2,y2) through which the line
should pass. These points are converted to give the usual equation for a line
\begin{equation}
y = mx + c.
\end{equation}
This line is then divided evenly into pixels to which the particles
may contribute. The contributions along this line from the particles is computed
as follows: 
\begin{figure}
\begin{center}
\includegraphics[width=0.8\columnwidth]{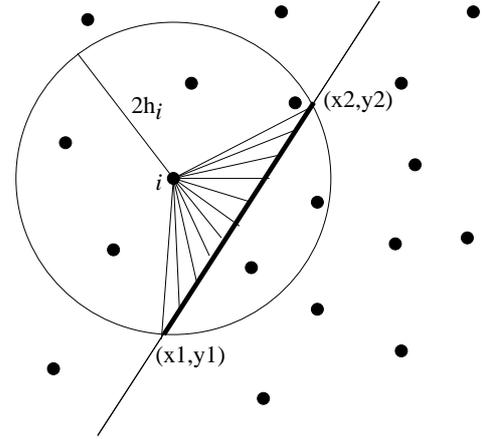}
\caption{Computation of a one dimensional cross section through 2D data. Each particle contributes to a sequence of pixels along the section of the cross-section line (if any) that intersects the smoothing circle.}
\label{fig:xsec2D}
\end{center}
\end{figure}
 For each particle, the points at which the cross section line intersects the
smoothing circle are calculated (illustrated in Figure \ref{fig:xsec2D}). The
smoothing circle of particle $i$ is defined by the equation
\begin{equation}
(x-x_i)^2 + (y-y_i)^2 = (2h)^2.
\end{equation}
The x-coordinates of the points of intersection are the solutions to the quadratic equation
\begin{equation}
(1 + m^2) x^2 + 2 (m (c - y_i) - x_i) x + (x_i^2 + y_i^2 - 2cy_i + c^2 - (2h)^2)= 0.
\end{equation}
For particles which do not contribute to the cross section line, the determinant
is negative. For the particles that do, it is then a simple matter of looping
over the pixels which lie between the two points of intersection, calculating
the contribution to each pixel using the 1D SPH summation interpolant, ie.
\begin{equation}
A(x) = \sum_j w_j A_{j} \mathcal{W}(\vert x - x_j \vert / h_j).
\end{equation}
 \begin{figure}
\begin{center}
\includegraphics[height=\columnwidth,angle=270]{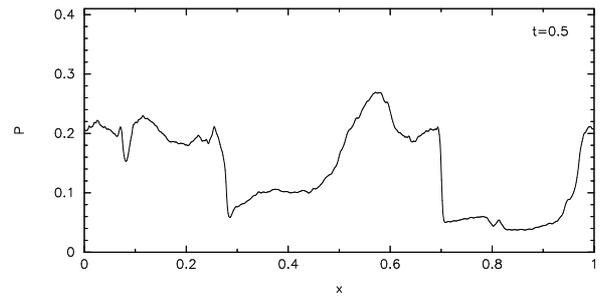}
\caption{Example of a one dimensional cross-section through 2D data, in this case showing the pressure distribution along a $y=0.3125$ cut through a high resolution version of the simulation shown in Figure~\ref{fig:orstang}.}
\label{fig:xsec2Dexample}
\end{center}
\end{figure}

 An example of a 1D cross section through 2D data is shown in Figure
\ref{fig:xsec2Dexample}. In principle a similar method could be used for oblique cross sections
through 3D data. In this case we would need to find the intersection
between the smoothing sphere and the cross section plane. However
in 3D it is simpler just to rotate the particles first and then take
a straight cross section as described above.

\subsubsection{2D vector plots}
 Vector plots of 2D data are produced by interpolating the $x-$ and $y-$ components of the vector separately to the pixel array, which are then used to plot an array of arrows centred on the pixels, with length proportional to the vector magnitude. Each component is interpolated exactly as for scalar 2D data, ie. 
 \begin{eqnarray}
A_{x}(x,y) & = & \sum_j w_{j} A_{x,j} \mathcal{W}(r / h), \\ 
A_{y}(x,y) & = & \sum_j w_{j} A_{y,j} \mathcal{W}(r / h), \\ 
r & = & \sqrt{(x - x_j)^{2} + (y -y_j)^{2}},  \\
h & = & {\rm max}(h_{j}, \Delta/2). \label{eq:hmax}
\end{eqnarray}
The main difference between interpolation for vector plots and that used for rendered plots is that far fewer pixels are used for the arrow plots (otherwise arrows become indistinguishable). Thus in general the interpolation for vector plots is more like a smoothing procedure rather than an interpolation (ie. there are far more particles than pixels). Since we only calculate distances to the centres of pixel cells, this is where the minimum smoothing length given by (\ref{eq:hmax}) becomes particularly important in providing a smooth representation of the data. An example of a 2D vector plot is shown in the lower panel of Figure~\ref{fig:orstang}.

\subsubsection{Streamlines}
\label{sec:streamlines}
 For a two dimensional vector map, streamlines (``fieldlines'') of the vector field can be plotted by integrating the vector field to find the stream function, contours of which provide the field lines. The stream function is given by
\begin{equation}
\Phi(x,y) = \int v_{x}(x,y) {\rm dy} - \int v_{y}(x,y){\rm dx},
\end{equation}
such that
\begin{eqnarray}
v_{x} & = & \frac{\partial \Phi}{\partial y}, \\
v_{y} & = & -\frac{\partial \Phi}{\partial x}.
\end{eqnarray}
 In SPLASH we compute the integral based on the \emph{interpolated} velocity field on the pixel array using a simple trapezoidal-rule integration. As presently implemented, this procedure works quite well when the vector field is smooth but performs poorly where there are strong gradients present.

\subsection{Rendering of 3D data}
 In three dimensions we must take either a projection through the whole domain or a cross section slice.
 
\subsubsection{Projections (line of sight integration)}
\label{sec:proj}
 In the projection case we wish to obtain an integral of the rendered quantity along the line of sight. We
begin with the 3D SPH summation interpolant in the form
\begin{equation}
A(x,y,z) = \sum_j m_j \frac{A_j}{\rho_j} W(x - x_j, y-y_j, z-z_j, h_j)
\end{equation}
where $W$ is the usual (3D) cubic spline kernel (\ref{eq:cubicspline}). Taking the integral of both sides
along the line of sight (assumed to be along the z axis) we have
\begin{equation}
\int A(x,y,z) {\rm dz} = \sum_j m_j \frac{A_j}{\rho_j} \int W(x - x_j, y-y_j, z-z_j, h_j) {\rm dz}.
\end{equation}
This shows that the line-of-sight integration for three dimensions can be written as a two dimensional
interpolation
\begin{equation}
\mathcal{A}(x,y) = \int A(x,y,z) {\rm dz} = \sum_j m_j \frac{A_j}{\rho_j} Y(x - x_j, y-y_j, h_j).
\end{equation}
where the 2D kernel (denoted $Y$) is the 3D kernel integrated through one spatial
dimension, ie.
\begin{equation}
Y(x,y) = \int W(x,y,z) dz.
\end{equation}
For practical purposes we write $Y$ in the form
\begin{equation}
Y(r_{xy}, h) = \frac{1}{h^2} F(q_{xy})
\end{equation}
where $q_{xy} = r_{xy}/h$ and $F(q_{xy})$ is the dimensionless 2D kernel given by
\begin{equation}
F(q_{xy}) = \int_{-\sqrt{R^{2} - q_{xy}^{2}}}^{\sqrt{R^{2} - q_{xy}^{2}}} \mathcal{W}(q){\rm dq_z} \label{eq:integratedform}
\end{equation}
where $q_z=z/h$, $q^{2}= q^{2}_{xy} + q^{2}_z$, $R$ is the kernel radius ($=2$ for the cubic spline) and $\mathcal{W}$ is the usual dimensionless kernel function for the cubic spline, ie.
\begin{equation}
\mathcal{W}(q) = \frac{1}{\pi} \left\{ \begin{array}{ll}
1 - \frac{3}{2}q^2 + \frac{3}{4}q^3, & 0 \le q < 1; \\
\frac{1}{4}(2-q)^3, & 1 \le q < 2; \\
0 & q \ge 2 \end{array} \right.
\end{equation}
The integral (\ref{eq:integratedform}) is not (obviously) tractable analytically (apart from at $q_{xy} = 0$). However it is
straightforward to perform this integration numerically (for all $q_{xy}$'s from $0\to 2$) and store the results in a table for the
interpolation calculation. This is the method adopted in SPLASH. An alternative would be to use a
different kernel in the visualisation for which the above integral can be calculated analytically.

 As previously, to avoid problems with round-off error we use the dimensionless weights defined in Equation (\ref{eq:weight}), thus writing the final interpolant (as implemented in the code) in the form
\begin{equation}
\mathcal{A}(x,y) = \int A {\rm dz} = \sum_j w_j h_{j} A_{j} F(r_{xy}/h),
\label{eq:3Dproj}
\end{equation}
where as previously we take
\begin{equation}
h = {\rm max}(h_{j}, \Delta/2).
\end{equation}
 An example of a 3D column-integrated plot is shown in Figure~\ref{fig:mcluster}, showing the results of a large scale star cluster formation calculation (in this case showing integrated density, ie. column density).

In the case of vector quantities each component is interpolated separately in the form
\begin{eqnarray}
\mathcal{A}_{x}(x,y) = \int A_{x} {\rm dz} & = & \sum_j w_j h_{j} A_{x,j} F(r_{xy}/h), \\
\mathcal{A}_{y}(x,y) = \int A_{y} {\rm dz} & = & \sum_j w_j h_{j} A_{y,j} F(r_{xy}/h), 
\end{eqnarray}
where again
\begin{equation}
h = {\rm max}(h_{j}, \Delta/2).
\end{equation}
This results in a line-of-sight integrated vector map which can be plotted on top of a rendered plot or as a standalone plot.

\begin{figure}
\begin{center}
\includegraphics[width=0.5\textwidth]{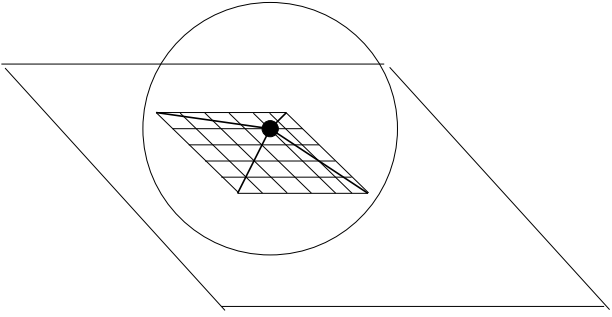}
\caption{Computation of a two dimensional cross section through 3D data: Each particle contributes to pixels in the cross section plane that lie within the smoothing sphere.}
\label{fig:xsec3D}
\end{center}
\end{figure}

\subsubsection{ Cross sections of 3D data}
 A cross section can be taken of three dimensional data by summing the
contributions to each pixel in the cross section plane from all particles within
$2h$ of the plane (Figure~\ref{fig:xsec3D}). In the implementation used in SPLASH the cross section is always at a fixed value of the third co-ordinate (ie. for xy plots the cross section is in the z direction). Oblique cross
sections can be taken by rotating the particles first (the combination of settings can be achieved easily in SPLASH's interactive mode by drawing a cross section plane with the mouse, from which the rotation angle and slice position are automatically calculated and the cross section subsequently plotted). The interpolation for cross sections (e.g. in $z$) takes the form
\begin{equation}
A(x,y,z_{0}) = \sum_j w_j A_j \mathcal{W}(r/h),
\end{equation}
where
\begin{eqnarray}
r & = & \sqrt{(x - x_{j})^{2} + (y - y_{j})^{2} + (z_{0} - z_{j})^{2}} \\
h & = & {\rm max}(h_{j}, \Delta/2),
\end{eqnarray}
and $z_{0}$ refers to the position of the cross section slice. As previously, vector plots are achieved by interpolating each component separately. Examples of 3D cross section plots are shown in Figure~\ref{fig:normalisation} of both scalar and vector fields.

\subsubsection{3D surface rendering of SPH data}
\label{sec:surface}
 A further option for visualisation of 3D data is to use surface rendering (see section \ref{sec:renderplots}). The idea is to produce a visualisation of
the surface of a data set by performing a ray-trace through the SPH particles, with the density
distribution giving the optical depth and the rendered quantity providing the colour. Thus
low-density regions will be transparent whilst high density regions will be opaque.

For a homogeneous medium the transport equation for a ray traced from $0\to D$ is
\begin{equation}
I_\nu(D) = I_\nu(0) e^{-\tau_\nu (D)} + S_\nu(1 - e^{-\tau_\nu(D)}),
\label{eq:transport}
\end{equation}
where $I_\nu$ is the (frequency dependent) intensity, $S_\nu$ is the source function along the ray
and $\tau$ is the monochromatic optical depth. The first term in (\ref{eq:transport}) represents absorption (intensity decreases by $e^{-\tau}$) whilst the second term represents emission. For example, at large
optical depth ($\tau\to\infty$) everything is obscured and all we see is the source function (ie.
light emitted from D), whereas at low optical depth $\tau\to 0$ the source function contributes
nothing and all we see is the previous intensity $I(0)$.

The optical depth $\tau$ is given by
\begin{equation}
\tau(D) = \int \kappa \rho {\rm ds},
\end{equation}
where $\rho$ is the density and $\kappa$ is the opacity (with dimensions of ``cross section per unit
mass").

 For SPH visualisation the procedure is as follows. First of all we sort the particles in `z' (where
$z$ represents the distance from the observer to the particle. Then starting from the furthest
particles, we consider the attenuation of a ray through each particle. Since what we are after is a
final 2D pixel map, what we do in practise is take one ray for each pixel, but rather than taking a
ray at a time (and looping over particles), we loop over all of the particles (from back to front),
calculating the contribution of that particle to all rays (`pixels') in the final pixel map. The
optical depth through the particle is given by
\begin{equation}
\tau(x,y) = \kappa \int \rho {\rm dz},
\end{equation}
where we have assumed that the opacity $\kappa$ is independent of $z$. Using the SPH summation for
the density, we have
\begin{equation}
\tau(x,y) = \kappa \sum_j m_j \int W(\vert {\bf r} - {\bf r}_j \vert,h) {\rm dz},
\label{eq:tauxy}
\end{equation}
giving just a summation involving the SPH kernel integrated through one spatial dimension, which is
the same as is used in the 3D projections (see $\ref{sec:proj}$ for details of how we compute this).
All that remains is to adjust $\kappa$ appropriately to give the desired surface position. In
SPLASH an approximate value for $\kappa$ is computed according to
\begin{equation}
\kappa = \frac{\pi \bar{h}^2 }{ (\bar{m}Y(0) d)}
\end{equation} 
where $\bar{h}$ and $\bar{m}$ are estimates for the average smoothing length and particle mass, calculated from the current (fixed) plot limits according to $\bar{h} = 0.5*(h_{min} + h_{max})$
(similarly for $\bar{m}$ - the important aspect here is that these values do not change between dump
files and can be restored from saved settings) and $Y(0)$ is the value of the integrated kernel function
(\S\ref{sec:proj})at the origin. The dimensionless parameter $d$ is then a user defined value giving
approximately the surface depth in terms of ``number of smoothing lengths".

 Actually, rather than computing the sum in Equation~(\ref{eq:tauxy}) for the whole ray, we consider just the attenuation of the ray through one particle at a time, using the optical depth for that particle alone.
Looping over each particle, we calculate the contribution to all rays (pixels) within the kernel radius $2h$. That is we have, for each particle
\begin{equation}
I(x,y) = I_{0}(x,y) e^{-\tau_i(x,y)} + S_i(1 - e^{-\tau_i(x,y)}),
\label{eq:transportp}
\end{equation} 
where $S_i$ is the source function (discussed below) and the optical depth through the particle's reach is
\begin{equation}
\tau_i (x,y) = \kappa m_i Y(x-x_i,y-y_i,h),
\end{equation}
where $Y$ is the integrated kernel function as in \S\ref{sec:proj}.

 In the computation of the surface rendering, there are two ways of proceeding. The first option is to assign each
particle an actual red, green and blue colour corresponding to the particle's value of the rendered
quantity (ie. from the colour table). The source function then consists of a red, green and blue
intensity $S_{i(r)},S_{i(g)},S_{i(b)}$. Then we would add up [ie. using (\ref{eq:transportp})] the intensities in each colour (red, green and blue) to get final red, green and blue values at each pixel. The effect of this is to ``blend" colours (so a red plus blue would make
purple), which is more like what happens in a real gas, but is meaningless in the sense that the
colours produced may no longer correspond to those in the colour table.
 
 The alternative is to use a `monochromatic' intensity - that is where the source function $S_i$ for each
particle is just the value of the rendered quantity at the particle location. Alongside this a
`total' optical depth is computed along each ray. Again, we add up the
intensities according to (\ref{eq:transport}), but now there is only a single value of $I$ for each
pixel, which corresponds to a final ``ray-averaged" value of the rendered quantity. The pixel map can
then be rendered in the usual manner using the ray-averaged values (which represent the values of the
rendered quantity at the `last scattering surface'). The only complication here is that we must make
the particles optically thin to the background. Thus the final colours must be faded to the
background colour (ie. black) according to the total optical depth computed for each pixel.
The latter method is the one used in SPLASH. However here we run into a limitation of the PGPLOT libraries, namely that the devices are limited to 256 colours, whereas we require 256 colours also at various degrees of blackness. Thus at present a non-faded version is returned to the PGPLOT device whilst a full (faded) version is written directly as a .ppm file (although without axes and annotation). This is one of the limitations that would make it desirable to change the back-end graphics library in future.

 An example of 3D surface rendering is shown in Figure~\ref{fig:nsns}, showing temperature in a simulation of the merger of two neutron stars.

\subsection{Rotation \& 3D perspective}
 Added perspective can be given to 3D plots by rotating the particles (``parallel projection'') or using a depth-dependent 3D perspective (that is, so that objects further away appear smaller). For SPH visualisation it is straightforward to apply these transformations to the particle positions prior to the interpolation procedure. The algorithm for 3D perspective is described below.
 
\subsubsection{ 3D perspective}
\begin{figure}
\begin{center}
\includegraphics[width=0.5\textwidth]{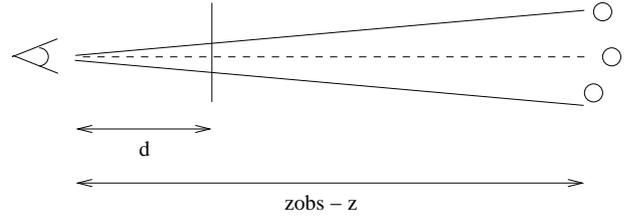}
\caption{3D perspective: Objects at a distance $d$ from the observer appear with unit magnification, whereas objects further away appear progressively smaller depending on their distance from the observer.}
\label{fig:3Dperspective}
\end{center}
\end{figure}

 3D perspective (illustrated in Figure~\ref{fig:3Dperspective}) is defined by two parameters: a distance to the observer (which we will call $z_{OBS}$) and a distance between the observer and a screen placed in front of the observer (which we will call $d$).
The transformation from usual $x$ and $y$ to screen $x'$ and $y'$ is then given by
\begin{eqnarray}
x' & = & x*d/(z_{OBS}-z), \nonumber \\
y' & = & y*d/(z_{OBS}-z).
\end{eqnarray}
 This means that objects at the distance $d$ will have unit magnification, objects closer than the
screen will appear larger (points diverge) and objects further away will appear smaller (points
converge). The SPLASH default is a $1/10$ reduction at the typical distance of the object (ie.
observer is placed at a distance of $10\times$ object size with distance to screen of $1\times$ object
size. SPLASH sets this as default using the current `z' plot limits as the `object size'.

 When using 3D perspective on interpolated plots the smoothing lengths of the particles are also modified by the 3D perspective, although the smoothing length used to give the $z$ length scale on integrated plots (Equation \ref{eq:3Dproj}) remains unchanged.
 
\section{Other useful techniques}
\label{sec:other}

\subsection{Fast particle plotting}
 Without using hardware graphics rendering, plotting large numbers of particles to the screen can be quite slow (certainly too slow for interactive work) and produces unnecessarily large files on vector plotting devices (e.g. postscript). Whilst one of the prime motivations behind SPLASH is to remove the need for raw particle plots as a poor man's SPH visualisation, plots showing correlations between certain variables or radial profiles can still require projected plots of large numbers of particles.
 
 SPLASH uses a simple trick to speed up this kind of plotting by dividing the plot surface into an array of pixels (typically $500 \times 500$) and plotting up to a maximum of 2 particles in each cell. This results in a substantial speed increase with almost no loss in visible information. Note that upon zooming the same criterion is applied to the zoomed-in view surface, so the effective resolution is increased appropriately.
 
\subsection{Accelerated rendering}
\label{sec:accel}
 The slowest of the rendering techniques is the calculation of a 3D projection through particles (\S\ref{sec:proj}) and the 3D surface rendering (\S\ref{sec:surface}) since they both involve contributions from all of the particles in the simulation, not just a subset. The former has the advantage that it can be easily parallelised (done so using OpenMP in SPLASH) whilst the latter is more complicated to implement in parallel (since for the surface rendering the contributions at each $z$ must be added in order). However a simple optimisation can be applied in both cases by taking advantage of the spherical symmetry of the kernel function.
 
  For example, considering the interpolation to the pixels shown in Figure~\ref{fig:2Dinterp} it is apparent that, provided we assume that the particle lies in the centre of the pixel which contains it, that the contribution to each quarter of the domain will be the same. Thus we can perform the interpolation to the top quarter of pixels only and copy the result to the remaining three quarters, providing an in-principle speedup of 4 for particles contributing to large numbers of pixels. The caveat is the assumption that the particle lies in the centre of the pixel. In practise the optimisation works well (that is, the results are visually identical to the non-optimised version) except where the particles are regularly distributed in the domain (e.g. on a lattice in the initial conditions), in which case the shift in the particle positions can produce unwanted grid patterns in the interpolation. For this reason the `accerated rendering' option is off by default but can be turned on by the user.

\section{Performance and memory usage}
\label{sec:perform}
 As discussed above, the slowest rendering techniques used in SPLASH are the calculation of a 3D projection through particles and the 3D surface rendering. However, even these are sufficiently fast to be performed interactively.  The algorithmic cost of the interpolation scales like $N_{part} \times N_{pix}^{2}$, where $N_{pix}$ is the number of pixels to which each particular particle contributes. Thus larger images are more expensive. In SPLASH the default number of pixels is set quite low (ie. $200\times 200$), with the idea being that a smaller number of pixels can be used for interactive work with the final step in producing the finished image to use a larger number of pixels.
  
 Whilst it is difficult to give precise timings (because the exact time taken for the rendering depends, amongst other things, on how many pixels each particle contributes to and thus how clustered the data is), SPLASH is easily able to handle very large data sets interactively in reasonable times. For example, producing a rendered projection of column density from a three dimensional simulation containing 135 million particles to a $600\times 600$ pixel image takes approximately 55 seconds on a single processor of our local supercomputer. Using the (shared-memory) parallel version on 8 cores of the same machine takes approximately 12 seconds. Similarly a 100 million particle simulation of a galactic disc takes approximately 26 seconds to render to a $1000\times1000$ pixel image on a single processor and around 7 seconds on 8 cores. Using the accelerated rendering technique described above (\S\ref{sec:accel}) results in a factor of 2-3 speedup on these timings. Surface rendering is somewhat slower -- approximately a factor of two more expensive than a column-integrated projection and currently not implemented in parallel. However the surface rendering technique is also not as widely applicable to different types of simulation.

 In terms of memory use, by default SPLASH reads into memory an entire dump file, converted to a two-dimensional single precision array (where the dimensions are the number of particles $\times$ the number of columns). Thus for a typical ``full dump file'' from a simulation of $10^{6}$ particles with 10 quantities ($x$, $y$, $z$, $v_{x}$, $v_{y}$, $v_{z}$, particle mass, smoothing length, density and thermal energy) this would require approximately 40Mb of storage (and hence 400Mb for $10^{7}$ particles, 4Gb for $10^{8}$ particles, etc.). Additionally a 4-byte integer colour index is stored for each particle and temporary memory is allocated for the two dimensional pixel array which is rendered to the screen, the size of which depends on the number of pixels chosen by the user (for example a $1000\times 1000$ image would require a further 4Mb). A low-memory mode for large datasets where memory is only allocated for those columns actually required to make a particular plot is currently being implemented (though applicable only to binary formats where data columns can be read independently). In this mode the data is re-read from disk every time a different plot is made (e.g. when plotting a $z-x$ projection of column density instead of an $x-y$ projection). Also plotting functionality which requires additional storage (such as particle colouring) will eventually be disabled in this mode.
 
  For smaller datasets, SPLASH can also be set to ``buffer'' all of the dumpfiles into memory, thus with memory requirements similar to the above times the number of dump files buffered. This provides a faster visualisation across multiple files for small datasets (since data does not have to be read from disk), provided sufficient memory is available. 

  For applications involving of the order of $10^{6}$ particles (typical of many current SPH simulations), the slowest part of movie-making (ie. applying the same visualisation to a series of data files) is reading the data from disk. To speed up the visualisation in this case SPLASH flags whether or not each particular column is required for the image being produced. For data reads where columns can be read independently (including that for the GADGET code) this is then used to read only the required subsection of the data from the dump file, resulting in a much faster data read.

\section{Summary/roadmap}
\label{sec:summary}
 In this paper we have presented SPLASH, a software tool for the visualisation of data from astrophysical Smoothed Particle Hydrodynamics simulations. The program is fully interactive, reads data direct from code dumps and can be used to visualise both scalar and vector SPH data in 1, 2 and 3 dimensions both to the screen and also to a variety of plotting devices provided by the PGPLOT graphics library. The software is designed to provide the user with a rapid feel for the output of a simulation and a variety of efficiently-implemented visualisation techniques unique to SPH with which to represent the results. There are many other features of SPLASH not discussed in this paper which we leave the reader to discover for themselves (and are described in the SPLASH userguide). These include:
\begin{itemize}
\item setting of animation sequences between frames in a movie
\item exact solutions to common test problems (e.g. hydrodynamic shock tubes, sedov blast wave)
\item transformation to different coordinate systems (e.g. cylindrical, spherical and toroidal coordinates)
\item particle tracking limits
\item multiple plots per page (appropriately tiled where so desired)
\item calculation of quantities not dumped
\end{itemize}

 Development of and improvements to the algorithms in SPLASH continues apace, particularly as a result of user feedback which has already helped to improve certain aspects of the program substantially. In terms of future developments, most notable is the absence of a routine for plotting stream/field lines through 3D SPH data and it would be highly desirable to be able to do this efficiently directly from the particles (rather than highly inefficiently via interpolation to a 3D grid).

  Secondly in several places SPLASH has outgrown the capacities of the PGPLOT graphics library. There are now several other graphics libraries layered on similar interfaces to PGPLOT (e.g. PLPLOT and s2plot, \citealt{s2plot}) and migrating the back-end library to one of these would not represent a formidable challenge. A more challenging alternative would be to move directly to OpenGL rendering, primarily for the speedup but also for the ease with which complicated 3D graphics can be manipulated. However the difficulty with at least the last two of these (OpenGL and s2plot) at present is that inherent in the SPLASH design is \emph{also} the ability to produce, non-interactively, appropriately annotated graphics for use in research papers. Similarly the visualisation should apply as easily to a series of dump files (non-interactively) as it does to a single file (interactively or not).

 In summary, SPLASH is an efficient and capable software package which makes the visualisation of SPH data a straightforward and enjoyable task for the user.  SPLASH is publicly available from
\url{http://www.astro.ex.ac.uk/people/dprice/splash/} and is released under the terms of the Gnu General Public Licence.

\section*{Acknowledgments} %If needed
 My knowledge of the techniques used for SPH visualisation presented here owes a great deal to many useful discussions with Matthew Bate. Useful conversations with Richard West and Klaus Dolag have also contributed to my understanding of the surface rendering technique. My research at the University of Exeter is supported by a UK PPARC/STFC postdoctoral research fellowship. Thanks also to the many users who have given feedback which has helped to improve SPLASH substantially and to the referee(s) for useful suggestions on this paper.

\bibliography{sph,starformation}

%\end{multicols}

\end{document}